\documentclass[%
 reprint,
 nobibnotes,
 amsmath,
 amssymb,
 aps,
 prl,
 groupaddress,
 floatfix,
 longbibliography,
 noeprint
 ]{revtex4-2}
\usepackage{multirow}
\usepackage{orcidlink}
\usepackage{graphicx}
\usepackage{float}
\usepackage{xcolor}
\usepackage{hyperref}\hypersetup{hidelinks,colorlinks=false}
\usepackage{placeins}
\usepackage{silence}

\begin{document}

\begin{flushright}{\large PITT-PACC-2618-v1}\end{flushright}

\title{Equiprobability Anomaly Detector Using Angular Distributions}

\author{Alex Whitten\,\orcidlink{0009-0007-8161-2904}}
\author{John Cicero\,\orcidlink{0009-0005-1191-2923}}
\author{Tae Min Hong\,\orcidlink{0000-0001-7834-328X}}
\email{tmhong@pitt.edu}
\affiliation{PITT PACC and Department of Physics and Astronomy, University of Pittsburgh, Pittsburgh PA, 15260
\vspace{6pt}}
\date{\today}

\begin{abstract}
\noindent
We use the angular distribution of top quark pairs ($t\bar{t}$) in the semileptonic final state $b\bar{b}\,\ell\nu q\bar{q}'$ to construct an anomaly detector for non-$t\bar{t}$ events.
Wigner $D$ functions expand the distribution with coefficients determined by orthogonal series density estimation.
Its equiprobable surfaces bin the events, the non-$t\bar{t}$ concentrate at low $t\bar{t}$ density.
The separation is quantified by the Kullback-Leibler divergence, which makes connection to variational autoencoders.
The performance is comparable to normalizing flows at parton level, but is limited by jet assignments at reconstruction level.
\end{abstract}

\maketitle

\textit{Introduction}---The probability density of an interaction $i$ is often a function of $n$ kinematic variables,
$
    f_i{:}~\mathbb{R}^n{\,\rightarrow\,}\mathbb{R}.
$
We write $\mathbf{x}$ to be the collection of such quantities characterizing an event, so the probability density is evaluated to return a single numerical value,
$
    y_i{\,=\,}f_i(\mathbf{x}).
$
It is normalized, i.e.,
$
    \int f_i(\mathbf{x})\,\mathrm{d}x_1\cdots\mathrm{d}x_n{\,=\,}1,
$
and
$
    N_i\,f_i(\mathbf{x})
$
is the number density.
For some phase space dimension $m{\,\le\,}n$, the $m$-fold differential distribution is
$
    N_i\,f_i(\mathbf{x}){\,=\,}%
   \partial^{m}{N_i}\big/ \partial x_1\cdots \partial x_{m}.
$
In 1d, $\mathrm{d}N/\mathrm{d}x\big|_{x_b}{\cdot\,}\Delta x$ is the number of events at $x_b$ in bin $b$ of width $\Delta x$.
In this paper we consider a differential distribution with respect to the value of the probability density $y_i$, $\mathrm{d}N/\mathrm{d}y_i$.
With $f_i$, we can compute $y_i$ for a different sample $j{\,\ne\,}i$ and its distribution would be higher near zero than that for $i$, because the number of $i$ events at a value $y_i$ is proportional to $y_i$ by definition.
So, unless $f_j$, whose distribution we do not consider explicitly, coincidentally vanishes where $f_i$ does,
\begin{align}
    \frac{1}{N_i}\frac{\mathrm{d}{N_i}}{\mathrm{d}{y_i}} \ll 
    \frac{1}{N_j}\frac{\mathrm{d}{N_j}}{\mathrm{d}{y_i}}
    \textrm{ as }y_i\rightarrow 0,
    \label{eq:grows}
\end{align}
which we use to construct an anomaly detector.

We consider top quark pairs ($t\bar{t}$)~\cite{Bernreuther:2004jv,Mahlon:2010gw,Czakon:2013goa} in the semileptonic final state $b\bar{b}\,\ell\nu{q}\bar{q}'$ as process $i$, described by angular variables to illustrate the method, and two processes $j$ as anomalies: standard model (SM) Higgs pairs  \cite{DiMicco:2019ngk,Grazzini:2018bsd,CMS:2026nuu,Papaefstathiou:2012qe} $hh{\,\rightarrow\,}b\bar{b}WW^*$  \cite{ATLAS:2018fpd,CMS:2024rgy} and, as a benchmark beyond the SM (BSM), heavy Higgs pairs $H\!H{\,\rightarrow\,}b\bar{b}WW$ with $m_H{\,=\,}170\,$GeV~\cite{Glover:1987nx,Plehn:1996wb,Dawson:1998py}.
Discriminants for $t\bar{t}$-vs-$hh$ classification \cite{Graesser:2012qy,Kim:2018cxf,Kim:2019wns,Alves:2025slf} exist, but the anomaly approach does not need a signal model and can be generalized to SM-vs-not(SM)~\cite{DAgnolo:2018cun,DAgnolo:2019vbw}.
However, as discussed later, the $W^*$ sculpts the angles, so more attention is given on $H\!H$.

\begin{figure}[b!]
    \centering
    \includegraphics[width=1.0\linewidth]{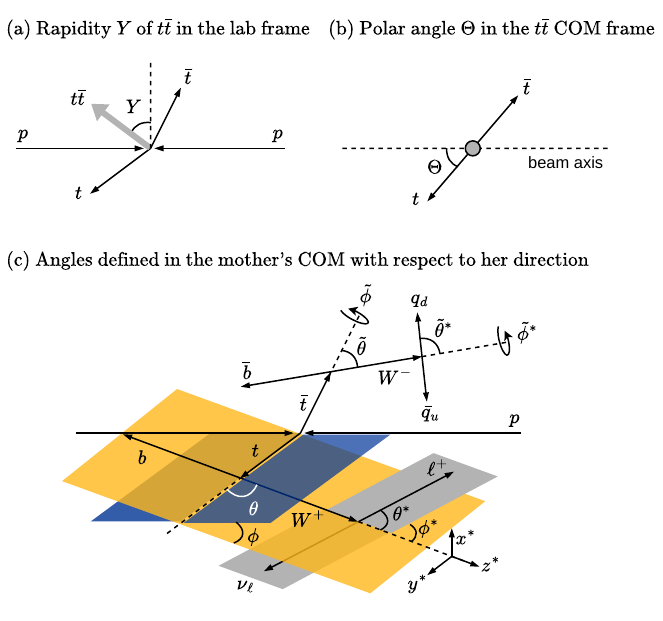}
    \vspace{-18pt}%
    \caption{Definition of the ten angles: (a) rapidity $Y$ of the $t\bar{t}$ system, (b) polar angle $\Theta$ of the leptonic top in the $t\bar{t}$ COM frame, (c) eight angles for the two top quark decays. Decay planes are shown only for the leptonic side for simplicity.}
    \label{fig:angledef}
\end{figure}

The semileptonic final state defines angles as shown in Fig.\ \ref{fig:angledef}.
We adopt the techniques of the studies of decays of a top quark in $Wtb$ interactions, which developed a robust formalism for angular distributions in order to probe BSM effects in the context of effective field theory~\cite{Kane:1991bg,Aguilar-Saavedra:2010ljg,Boudreau:2013yna,Aguilar-Saavedra:2015yza,Aguilar-Saavedra:2017wpl}.
In particular, its use of orthogonal series density estimation (OSDE) to extract the coefficients using a finite sample is adopted~\cite{Boudreau:2016pdi,ATLAS:2015ryj,ATLAS:2017rcx,ATLAS:2025adk}.

In the literature, anomaly detectors \cite{Kasieczka:2021xcg,Aarrestad:2021oeb,Karagiorgi:2021ngt,Belis:2023mqs,Caron:2024zxk} use machine learning (ML) \cite[e.g.,][]{Nachman:2020lpy,Hallin:2021wme,Stein:2020rou,Liu:2012,CrispimRomao:2020ojx,Chebrolu:2026,Caron:2021wmq,Buss:2022lxw,Metodiev:2023izu,Das:2026ysm,Dillon:2022mkq,Krause:2023uww} to learn the information density, such as normalizing flows or autoencoders (AE)~\cite{Cerri:2018anq,Farina:2018fyg,Heimel:2018mkt,Govorkova:2021utb,Roche:2023int}, and are deployed at the LHC~\cite{ATLAS:2020iwa,ATLAS:2023ixc,CMS:2024nsz}.
A more physics-based approach is to rank events by $|\mathcal{M}|^2$ \cite{Debnath:2014eaa} or to assign final-state particles to partons~\cite{Dillon:2025dxr}.
This paper follows the $|\mathcal{M}|^2$ method, but only impose angular momentum conservation.
We connect the information-theoretic aspects of the distribution of $y$ to the volume $V$ of equiprobable surfaces via the identity $\mathrm{d}N/\mathrm{d}y{\,=\,}N\,\mathrm{d}V/\mathrm{d}({-}\ln y)$ and to the variational AE \cite{Kingma:2013hel,Rezende:2014} via the Kullback-Leibler divergence \cite{Kullback:1951}.

\textit{Setup}---A sample of $10^5$ $t\bar{t}$ events is generated using the default SM settings, with an independent sample of $10^4$ events for testing;
$10^5$ $H\!H$ and $2{\,\cdot\,}10^4$ $hh$ events using the HEFT package with NNPDF2.3~\cite{Ball:2012cx} at LO.
The samples are generated by MadGraph5\_aMC@NLO 3.6.7 \cite{Alwall:2014hca} at leading order for proton collisions at $\sqrt{s}{\,=\,}13\,$TeV.
For the reconstruction-level study, the events are showered and hadronized with Pythia~8.3~\cite{Bierlich:2022pfr} and passed through Delphes~3~\cite{deFavereau:2013fsa} with its ATLAS card without pileup.
Jets are reconstructed from calorimeter towers with the anti-$k_t$ algorithm~\cite{Cacciari:2008gp} with $R{\,=\,}0.6$ using FastJet~\cite{Cacciari:2011ma}, and are required to have $p_\mathrm{T}{\,>\,}25\,$GeV and $|\eta|{\,<\,}2.5$; $b$ jets are identified with the parametrized efficiencies of Ref.~\cite{ATL-PHYS-PUB-2015-022}.
Leptons are required to have $p_\mathrm{T}{\,>\,}25\,$GeV and $|\eta|{\,<\,}2.47$ ($2.5$) for electrons (muons) to mimic the trigger selection.
Events are selected with exactly one lepton, at least four jets and at least one $b$ jet.
The $\nu$ transverse momentum is taken from $E_\mathrm{T}^\mathrm{miss}$, and its longitudinal momentum from the $m_W$ constraint, which has two solutions.
The four jets and the $\nu$ solution are assigned to the decay products by minimizing
\begin{align}
    \!\chi^2{\,=}\left(\frac{m_{jj}-m_W}{\sigma_W}\!\right)^{\!2}%
    {\!\!+\!}\left(\frac{m_{jjb}-m_t}{\sigma_t}\!\right)^{\!2}%
    {\!\!+\!}\left(\frac{m_{\ell\nu b}-m_t}{\sigma_t}\!\right)^{\!2}\!
\end{align}
over the permutations of the six leading jets, with up to two $b$ jets, taking $\sigma_W{\,=\,}10\,$GeV and $\sigma_t{\,=\,}15\,$GeV.
The selection is passed by 24\% (20\%) of the $t\bar{t}$ ($H\!H$) events.
To isolate the effect of the jet assignment, the reconstructed jets are also matched one-to-one to the four quarks within $\Delta R{\,<\,}0.4$, each jet taking the role of its quark, with the $\nu$ solution whose $m_{\ell\nu b}$ is closer to $m_t$; 7.1\% (3.5\%) of the $t\bar{t}$ ($H\!H$) events pass the selection and are matched.

\textit{Angular Distributions}---We introduce the notation for OSDE~\cite{Cencov:1962,Schwartz:1967,Efromovich:2010}. For a given probability of a random variable $X$ over domain $R$, $P(X){\,=}\int_R f(x)\,\textrm{d}x$, the probability density $f(x)$ can be expressed by a generalized Fourier series as
$
    \sum_i c_i\,\varphi_i(x),
$
where $\{\varphi\}$ is a complete set of orthonormal basis functions.
The coefficients
$
    c_i{\,=}\int_R \varphi_i^*(x)\,f(x)\,\textrm{d}x,
$
can be estimated, denoted by a hat over the quantity, by OSDE with a finite sample of $N$ events by the sample mean estimator~\cite{Beaujean:2015xea,Dighe:1998vk,Kronmal:1968},
$
    \hat{c}_i{\,=\,}N^{-1}\sum_{k=1}^N \varphi_i^*(x_k),
$
where $x_k$ is the $X$ value for the $k^\textrm{th}$ event and $\hat{c}_i{\,\to\,}c_i$ as $N{\,\to\,}\infty$.

To create an orthogonal basis on which to fit our discriminator we characterize the decay $t{\,\to\,}Wb$ followed by $W{\to\,}\ell\,\nu_\ell$ by four angles. The first pair ($\theta, \phi$) are the polar and azimuthal coordinates, respectively, of the $W$ boson momentum in the top quark rest frame, with respect to the plane defined by the beam axis and the top quark momentum in the lab frame.
The second pair ($\theta', \phi'$) are the spherical coordinates of the charged lepton momentum in the $W$ boson rest frame, in the reference system ($x', y', z'$) obtained by the standard boost from the previous one, such that the $z'$ axis is in the direction of the $W$ boson momentum and the $y'$ axis is in the $xy$ plane.
The angles $\boldsymbol{\theta}{\,=\,}\{\theta, \phi, \theta', \phi'\}$ are illustrated in Fig.~\ref{fig:angledef}.

The amplitude for the decay of the top quark $t{\,\to\,}Wb$ followed by its leptonic decay $W{\,\to\,}\ell\,\bar{\nu}_\ell$ can be written, in the narrow width approximation, by a set of $M$ functions
$
   M_{m'm}^{j_1j_2}(\phi,\theta,\phi',\theta'),
$
which is an orthonormal set for $j_1j_2m'm$
(see Appendix).
The differential cross section, at parton level, then, is written as
\begin{align}
\underbrace{\frac{1}{\Gamma}
\frac{\partial\Gamma}{\partial\Omega\,\partial\Omega'}(\boldsymbol{\theta})}_{f_t(\boldsymbol{\theta})}
=\sum_{j_1j_2m'm}^{j_1\le 1,\,j_2\le 2}c_{m'm}^{j_1 j_2}\cdot M_{m'm}^{j_1 j_2}(\boldsymbol{\theta}),
\label{eq:1top}
\end{align}
where $\Gamma$ is the decay width.
This is our function $f_t$ for single top.
There are 24 coefficients $c_{m'm}^{j_1j_2}$ that are combinations of $a_{\lambda_1\lambda_2}$ and the various polarizations and helicity fractions~\cite{Aguilar-Saavedra:2024vpd,Durupt:2025wuk,Bernreuther:2004jv,Mahlon:2010gw,CMS:2019nrx}, but whose exact expressions are not relevant, so we omit them.
More compactly, the set of indices $j_1j_2m'm$ is shortened to $\boldsymbol{j}$ for below.

The decay of the other top quark adds four angles $\tilde{\boldsymbol{\theta}}{\,=\,}\{\tilde{\theta},\tilde{\phi},\tilde{\theta}',\tilde{\phi}'\}$, defined as above with the down-type quark in place of the charged lepton.
The remaining two of the ten variables are the polar angle $\Theta$ of the leptonic top quark in the $t\bar{t}$ rest frame and the rapidity $Y$ of the $t\bar{t}$ system, whose distributions we expand empirically to five terms in normalized Legendre ($P_l$) and Hermite ($H_n$) polynomials \cite{Boudreau:priv}.
With $\mathbf{x}{\,=\,}\{\Theta,Y,\boldsymbol{\theta},\tilde{\boldsymbol{\theta}}\}$, Eqn.~\ref{eq:1top} becomes
\begin{align}
    f_{t\bar{t}}(\mathbf{x}){\,=}\sum\limits_{\boldsymbol{j\tilde{\jmath}}\,ln}c_{\boldsymbol{j\tilde{\jmath}}\,ln}{\,\cdot\,}
    \underbrace{
    M_{\boldsymbol{j}}(\boldsymbol{\theta})\,M_{\boldsymbol{\tilde{\jmath}}}(\tilde{\boldsymbol{\theta}})\,P_l(\cos\Theta)\,H_n(Y)
    }_{B_{\boldsymbol{j\tilde{\jmath}}\,ln}(\mathbf{x})},
    \label{eq:2top}
\end{align}
where $\boldsymbol{\tilde{\jmath}}$ indexes the hadronic decay and the $B$ are orthonormal by construction.
The terms with $j_1{\,=\,}\tilde{j}_1{\,=\,}1$ contain the $t\bar{t}$ spin correlations~\cite{Bernreuther:2015yna}, in which entanglement has been observed~\cite{ATLAS:2023fsd,CMS:2024zkc}.
The $24^2{\,\cdot\,}5^2{\,=\,}14400$ coefficients with $\mathrm{d}\mathbf{x}{\,=\,}\mathrm{d}\Omega\,\mathrm{d}\Omega'\,\mathrm{d}\tilde\Omega\,\mathrm{d}\tilde\Omega'\,\mathrm{d}\cos\Theta\,\mathrm{d}Y$,
\begin{align}
    c_{\boldsymbol{j\tilde{\jmath}}\,ln}=\int
    B_{\boldsymbol{j\tilde{\jmath}}\,ln}^{\,*}(\mathbf{x}){~}
    f_{t\bar{t}}(\mathbf{x}){~}
    \underbrace{
    \mathrm{d}x_1\cdots\mathrm{d}x_n
    }_{\mathrm{d}\mathbf{x}},
\end{align}
are estimated by the sample means of $B^*(\mathbf{x}_k)$.
Most of the estimates $\hat{c}_{\boldsymbol{j\tilde{\jmath}}\,ln}$ are consistent with zero, so we keep around 300 that are above 2.5 times the statistical noise whose threshold is found by a cross-validation study and evaluated on the non-trained slices.

\begin{figure}[t!]
    \centering
    \includegraphics[width=0.893\linewidth]{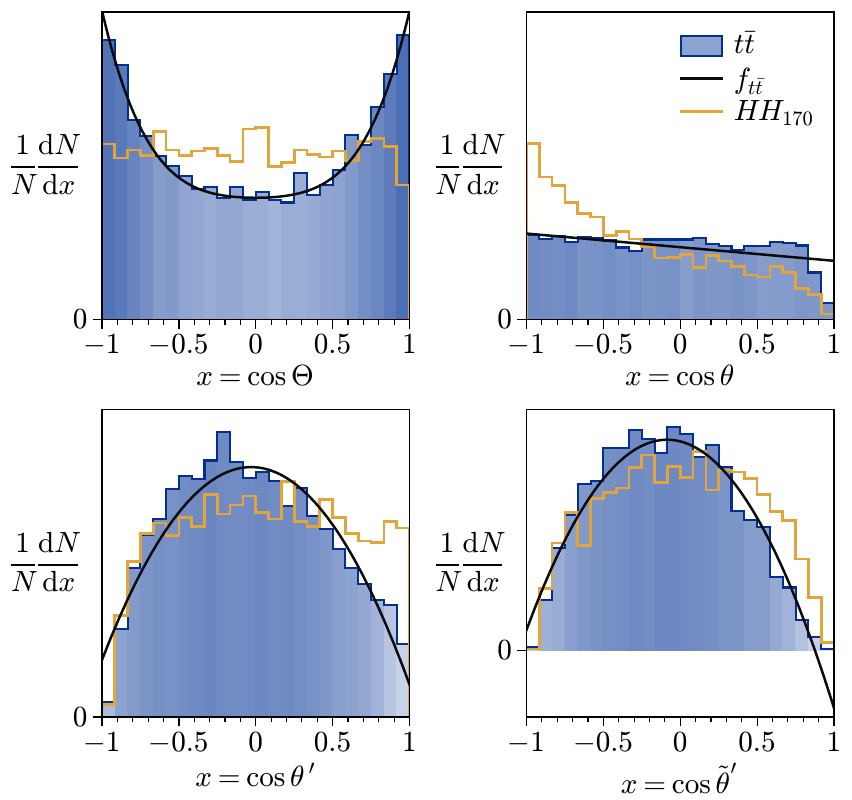}
    \vspace{-12pt}%
    \caption{Projections of the OSDE fit $f_{t\bar{t}}$ (line) compared with the $t\bar{t}$ fit sample (filled) for four of the ten angles.
    Reconstructed jets are matched to quarks within $\Delta R{\,<\,}0.4$.
    The $H\!H$ ($m_H{\,=\,}170\,$GeV) sample is overlaid.
    }
    \label{fig:OSDE}
\end{figure}

The distributions of the ten angles for the $t\bar{t}$ sample at reconstruction level with matched jets (see \emph{Setup}) agree with the results of the OSDE fits, integrated over the other nine angles, as shown for four of the angles in Fig.~\ref{fig:OSDE}. 
The fits are mostly well described within the plotted variable's domain of validity $R$, but plunge to negative values near the edges of $R$ due to the Gibbs overshoot, similarly to when describing a square pulse with a finite set of sinusoidal functions~\cite{Hewitt:1979,Gajek:1986} estimated over a finite sample.

\begin{figure}[t!]
    \centering
    \includegraphics[width=\linewidth]{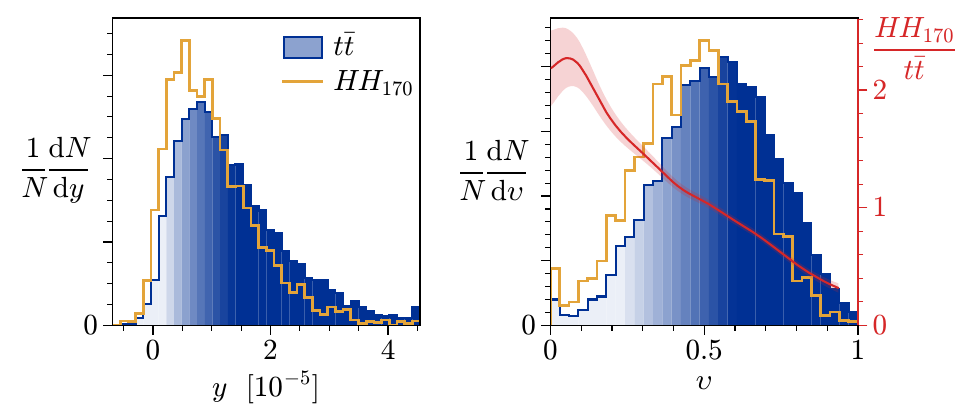}
    \vspace{-21pt}%
    \caption{Normalized distributions of the anomaly score for $t\bar{t}$ and $H\!H$ ($m_H{\,=\,}170\,$GeV) in the $b\bar{b}\,\ell\nu q\bar{q}'$ final state at the reconstruction level with the jets matched to the quarks within $\Delta R{\,<\,}0.4$.
    Left: $y{\,=\,}f_{t\bar{t}}(\mathbf{x})$ is the probability density.
    Right: $\upsilon$ is a transformation of $y$; it rescales and flips ${-}\ln{y}$ after shifting $y$ out of the negative region; details in \emph{Results}. The shading for the $t\bar{t}$ histogram matches the scale in Fig.\ \ref{fig:dNdy}.
    The red curve is the ratio $N_{H\!H}/N_{t\bar{t}}$ with its values on the right.
    \vspace{-6pt}%
    }
    \label{fig:p}
\end{figure}

\textit{Results}---The probability density function of Eqn.~\ref{eq:2top} is evaluated for the reconstructed samples and its results are shown in Fig.~\ref{fig:p}.
The left plot shows the distribution of $y{\,=\,}f_{t\bar{t}}(\mathbf{x})$, the value of the probability density.
The $y$ dips into the negative domain because of the Gibbs overshoot,
which contains $1\%$ ($3\%$) of the $t\bar{t}$ ($H\!H$) events (see Fig.~\ref{fig:p}).
We transform $y$ into $\upsilon$ as follows.
First, $y$ is shifted by $y_0$, the magnitude of its lowest value, in order to compute its logarithm, $y'{\,=\,}y{\,+\,}y_0$, so $s'{\,=\,}{-}\ln{y'}$.
Second, $s'$ is rescaled, flipped and normalized to define the discriminator $\upsilon$ to span a score from 0 to 1, with the former (latter) being less (more) like $t\bar{t}$ events, $\upsilon{\,=\,}(s'_1{\,-\,}s')/(s'_1{\,-\,}s'_0)$, where $s'_0$ ($s'_1$) is the $0.1\%$ ($99.5\%$) quantile of $s'$.
The right plot shows the distribution of $\upsilon$.

\begin{figure}[t!]
    \centering
    \includegraphics[width=0.65\linewidth]{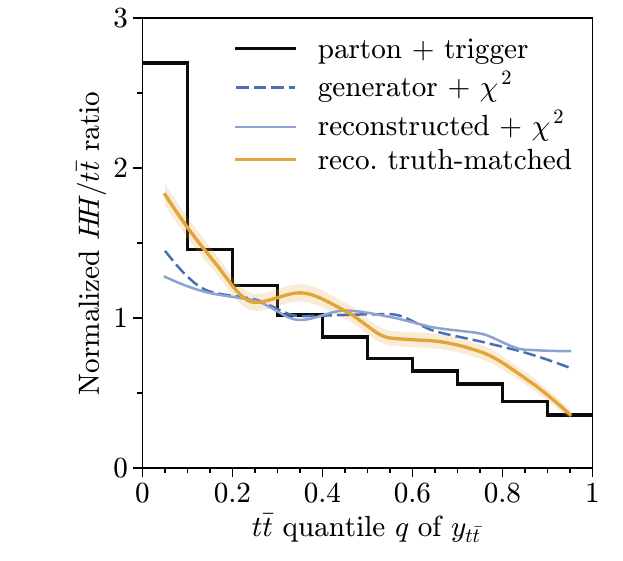}
    \vspace{-9pt}%
    \caption{Separation of $H\!H$ ($m_H{\,=\,}170\,$GeV) from $t\bar{t}$ by the $t\bar{t}$ density $y_{t\bar{t}}{\,=\,}f_{t\bar{t}}(\mathbf{x})$ at each level, with $f_{t\bar{t}}$ fit separately at each level.
    The ratio of the $H\!H$ and $t\bar{t}$ distributions is plotted versus the $t\bar{t}$ quantile $q$ of $y_{t\bar{t}}$.
    The curves are histograms in bins of width 0.1 in $q$ through the bin centers; the band on the truth-matched curve shows its statistical uncertainty.
    \vspace{-12pt}%
    }
    \label{fig:levels}
\end{figure}

The results are compared against two unsupervised and three supervised methods using the same ten angles $\mathbf{x}$.
For the equiprobability method of this paper, at parton level after the single-lepton trigger, the area under the ROC curve (AUC) against $t\bar{t}$ is 0.68/0.67 (for $hh$/$H\!H$ signals) and is comparable to 0.68/0.66 from a normalizing flow~\cite{Dinh:2016pgf} and better than 0.59/0.57 from an autoencoder~\cite{Farina:2018fyg,Heimel:2018mkt}, both evaluated on the independent test sample.
At the reconstruction level with the $\chi^2$ jet assignment,
correct for 19\% for $t\bar{t}$ and 7\% for $H\!H$, the AUC falls to 0.51/0.54.
With these jets matched to the quarks and $f_{t\bar{t}}$ refit, the $H\!H$ rises to 0.61 (see Fig.~\ref{fig:levels}).
The supervised classifiers (BDT~\cite{Friedman:2001,Pedregosa:2011}, MLP~\cite{Rumelhart:1986} and FT-Transformer~\cite{Gorishniy:2021}) trained with signal labels reach the higher value of AUC, as expected, all within 1\% of 0.89/0.92, and reduces to 0.63/0.69 after jet assignment.

\textit{Discussion}---The meaning of $\mathrm{d}{N}/\mathrm{d}{y}$ is discussed, followed by a list of limitations of our study.
Our approach is, in its core, the matrix element method~\cite{Kondo:1988yd,D0:2004rvt,Fiedler:2010sg,Artoisenet:2010cn,Gainer:2013iya}.
The probability density is $y{\,=\,}|\mathcal{M}|^2$ up to normalization, parton densities and transfer function~\cite{Note:optimal}.
The equiprobable contour is then the $(n{-}1)$d surface in $n$d phase space for equal values of $|\mathcal{M}|^2$, providing a distribution binned in $|\mathcal{M}|^2$.
In our formalism, the angles provide a natural measure \cite{Kasieczka:2022naq}, because $\mathrm{d}\Omega$ is rotation invariant, $\mathrm{d}Y$ is invariant under Lorentz boost along the beam direction and the basis functions are orthonormal, so $y$ is $|\mathcal{M}|^2$ without a Jacobian factor.

Physically, $hh$/$H\!H$ events are less well described, relative to $t\bar{t}$, by $f_{t\bar{t}}$ by construction.
Mathematically, $\mathrm{d}{N}/\mathrm{d}{y}$ is the change in the density of events (see Appendix):
\begin{align}
    \!\frac{1}{N}\frac{\mathrm{d}{N}}{\mathrm{d}{y}}
    \!{\,=\,}\frac{\mathrm{d}}{\mathrm{d}(\,\underbrace{{-}\ln{y}}_{s}\,\phantom{\Big|}\!\!\!)} \underbrace{\int_{f(\mathbf{x})>y}\!\!\!\!\!\!\!\!
    \mathrm{d}\mathbf{x}}_{V(y)}
    {\,=\,}y\underbrace{\oint_{f(\mathbf{x})=y}}_{S(y)}\!\!\!\!\!\!
    |\nabla f(\mathbf{x})|^{-1}\,\mathrm{d}\mathbf{a}.
    \!
    \label{eq:dNdy}
\end{align}%
Geometrically, $V(y)$ is the volume of phase space with $f(\mathbf{x}){\,>\,}y$ and $S(y)$ the equiprobable surface that bounds it; the second equality is the co-area formula~\cite{Federer:1959}.
In information-theoretic terms, $s_i(\mathbf{x}){\,=\,}{-}\ln f_i(\mathbf{x})$ is the surprisal, or Shannon information, of an event under process $i$.
Its mean value for process $j$,
\begin{align}
    \textstyle\langle s_i\rangle_j{\,=\,}\int f_j(\mathbf{x})\,s_i(\mathbf{x})\,\mathrm{d}\mathbf{x}
    {\,=\,}{-}\int f_j(\mathbf{x})\,\ln f_i(\mathbf{x})\,\mathrm{d}\mathbf{x},
\end{align}
is the cross-entropy of events $j$ scored with $f_i$ and reduces to the entropy for $i{\,=\,}j$~\cite{Shannon:1948,Note:Boltzmann,Nalisnick:2019}.
The cross-entropy exceeds the entropy by the Kullback-Leibler (KL) divergence, defined as
\begin{align}
D_\mathrm{KL}(f_j\,\|\,f_i){\,=\,}\langle s_i\rangle_j{\,-\,}\langle s_j\rangle_j{\,\ge\,}0,
\end{align}
a measure of how $j$ differs from $i$ that, unlike the probability density, does not depend on the choice of coordinates~\cite{Kasieczka:2022naq}.
Interestingly, $D_\mathrm{KL}$ appears in the loss function of the variational autoencoder (VAE), where it regularizes the encoded latent distribution toward its prior.
The VAE as anomaly detector at the LHC~\cite[e.g.,][]{Cerri:2018anq,Sugizaki:2025gel,Gonski:2026jgu,RealTimeML:2025,Govorkova:2021utb} is realized in physics analyses~\cite[e.g.,][]{ATLAS:2023dijetAD,CMS:2024dijetAD}, HLT and Level-1 trigger \cite{CMSECAL:2024,Zipper:2023ybp,Cohen:2026qug}.
Looking forward, building on the connections made between the equiprobability framework and VAE, a new ML design using the physics-based $D_\mathrm{KL}$ may offer built-in $|\mathcal{M}|^2$ constraints.

\begin{figure}[t!]
    \centering
    \includegraphics[width=0.95\linewidth]{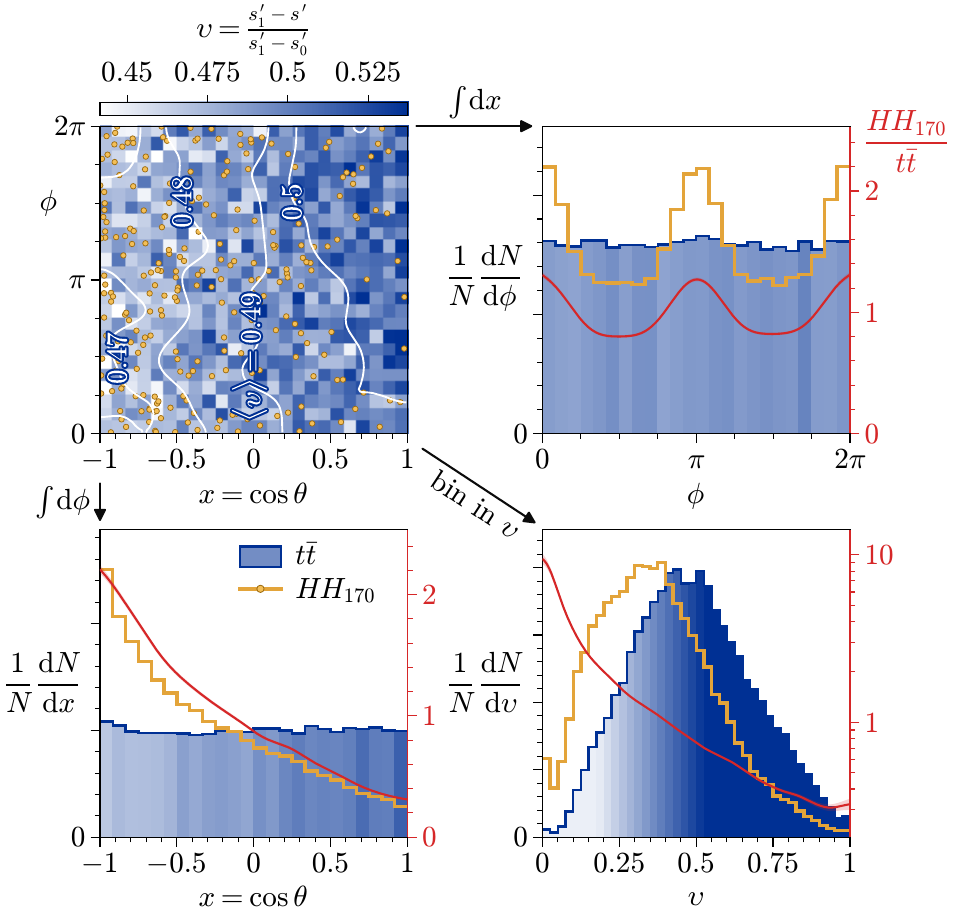}
    \vspace{-9pt}%
    \caption{
    Normalized differential distributions of $t\bar{t}$ and $H\!H$ ($m_H{\,=\,}170\,$GeV) samples at parton level after the single-lepton trigger, with $y$ and $\upsilon$ as in Fig.~\ref{fig:p}.
    The $\mathrm{d}N/\mathrm{d}\upsilon_{t\bar{t}}$ scatter plot integrates over the other eight variables so the contours are not closed as for the toy sample (see Fig.\ \ref{fig:toy}).
    The $\upsilon$ is a transformation of $y$, using $s'{\,=\,}{-}\ln{(y{\,+\,}y_0)}$ that shift the $y$ out of the negative region, to construct a discriminator.
    \vspace{-9pt}%
    }
    \label{fig:dNdy}
\end{figure}

For our $t\bar{t}$ study, the corresponding plots for the parton-level study are shown in Fig.\ \ref{fig:dNdy}.
As mentioned previously, we transform $y$ into $\upsilon$.
Because the scatter plot is a projection down from a 10d phase space, the equiprobable contours are not neatly closed as in the toy example (see Appendix).
What is evident is that the ratio $N_{H\!H}/N_{t\bar{t}}$, shown by the red curve with the numerical values on the right side of the plot, increases as $\upsilon{\,\rightarrow\,}0$.
The separation of $H\!H$ from $t\bar{t}$ is compared at parton level after the trigger, at generator and reconstruction level with the $\chi^2$ jet assignment and at reconstruction level with truth-matched jets (see Fig.~\ref{fig:levels}).
The corresponding AUC values are 0.67, 0.56, 0.54 and 0.61, respectively.
The probability density of $t\bar{t}$ is shown as quantile $q$ on the horizontal axis.
At every level the $H\!H$/$t\bar{t}$ ratio grows as $q{\,\rightarrow\,}0$.
In the lowest decile the ratio is 2.7 at parton level after trigger, 1.3 at reconstruction level with the $\chi^2$ jets and 1.8 with truth-matched jets.

The information content between the $t\bar{t}$ sample and the $H\!H$ sample can be extracted from their $y_{t\bar{t}}$ distributions.
Because $t\bar{t}$ is uniform in $q$, by construction, each curve in Fig.~\ref{fig:levels} represents the likelihood ratio $r_{H\!H}{\,=\,}f_{H\!H}/f_{t\bar{t}}$ averaged over the equiprobable surface at each value of $y_{t\bar{t}}$.
The difference between the two processes is quantified as the KL divergence that can be derived from the plot.
The ratio of densities in $q$ are $r_{H\!H}$ for $H\!H$ and 1 for $t\bar{t}$, so $D_\mathrm{KL}(r_{H\!H}\,\|\,1){\,=}\int_0^1 r_{H\!H}\ln r_{H\!H}\,\mathrm{d}q$, and the value is $0.21$ at parton level after the trigger, $0.022$ at generator level and $0.012{\,\pm\,}0.002$ at reconstruction level with the $\chi^2$ jet assignment.
As expected, the value is much higher at $0.079{\,\pm\,}0.009$ for truth-matched jets, where the errors are statistical.
The ratio of $D_\mathrm{KL}$ values helps quantify the relative gain: the reconstruction with $\chi^2$ jet assignment keeps $6\%{\,=\,}0.012/0.21$ of the parton-level information and the truth-matched jets $38\%{\,=\,}0.079/0.21$.

This study is a proof of principle and uses leading-order simulated samples without pileup or systematic uncertainties, with the $\nu$ momentum, jet assignment and quark flavors known at parton level.
For $hh$, the $m_W$ constraint on the neutrino $p_z$ forces the off-shell $W^*$ to look on-shell, so our angular method is not effective without further modification.
For $H\!H$, the $\chi^2$ jet assignment built on the $t\bar{t}$ masses forces the signal kinematics to look like $t\bar{t}$, washing out the parton-level separation (see Fig.~\ref{fig:levels}).
However, any assignment built on the $t\bar{t}$ hypothesis~\cite{Erdmann:2013rxa,Shmakov:2021qdz} would exhibit similar degradation.

\textit{Conclusions}---We presented a method to use the angular distribution of events of type $i$ to discriminate against another $j$ without any knowledge of the $j$'s distribution.
As an example, the distribution of simulated samples of $t\bar{t}$ was described using Wigner $D$ functions, whose coefficients were extracted using orthogonal series density estimation.
The $t\bar{t}$ probability density was then applied to non-$t\bar{t}$ samples as an anomaly detector.
In contrast to ML methods, equiprobable surfaces provide a physics-based approach to binning events.
Lastly, in relation to information theory, we connected our method with the KL divergence that appears in variational autoencoders.

\textit{Acknowledgments}---We thank Eli Ullman-Kissel for the initial work; Joe Boudreau for OSDE and Gibbs; Jim Mueller for basis functions; Brian Batell for information theory; Rui Xue for the angles.
TMH is supported by the US Department of Energy [award no.\ DE-SC0007914].

Claude AI (Anthropic) interpreted a complete draft of this paper with the $hh$ results to validate it and extended it to BSM scenarios.
It performed the coefficient study and the ML comparisons.
It aided with paper editing.

\textit{Code availability}---\href{https://github.com/tmhong/equiprobable}{github.com/tmhong/equiprobable}


\newpage
\appendix

\section{Appendix}
\vspace{-6pt}
\textit{Single top decay angular distribution}---The top quark decay in $Wtb$ can be described, at parton level, by
$
    A(\boldsymbol{\theta}){\,=\,}\sum_{\lambda_1}
    a_{\lambda_1\lambda_2}{\cdot\,}
    b_{\lambda_3\lambda_4}{\cdot\,}
    D^{1/2\,*}_{\mu\lambda}(\phi,\theta,0){\,\cdot\,}
    D^{1\,*}_{\lambda_1\lambda'}(\phi',\theta',0),
$
where $\lambda_{1\ldots4}$ represents the helicities of $W$, $b$, $\ell$ and $\bar{\nu}_\ell$, respectively and coefficients $a_{\lambda_1\lambda_2}$ and $b_{\lambda_3\lambda_4}$ weight the Wigner $D$ functions~\cite{Jacob:1959at}. The subscript $\mu$ represents the third spin component of the top quark, $\lambda{\,=\,}\lambda_1{\,-\,}\lambda_2$, $\lambda'{\,=\,}\lambda_3{\,-\,}\lambda_4$.
Angular momentum conservation keeps only four $a_{\lambda_1\lambda_2}$ terms for the top and one $b_{\lambda_3\lambda_4}$ term for the $W$, so we absorb $b_{\lambda_3\lambda_4}$ into the definitions of $a_{\lambda_1\lambda_2}$.
Following Ref.~\cite{Aguilar-Saavedra:2017wpl}, we define $M$ functions in Eqn.\ \ref{eq:1top} to collect the results of squaring the amplitude, condensing two $D$ functions into one using the composition rule and converting $D^*$ to $D$ by manipulating the indices.
\vspace{3pt}

\textit{Differential distribution of probability density}---We derive Eqn.~\ref{eq:dNdy} by writing the distribution of $N$ events in two ways, distinguishing the function $f$ from its value $y{\,=\,}f(\mathbf{x})$.
Inserting $\int\delta(f(\mathbf{x}){\,-\,}y)\,\mathrm{d}y{\,=\,}1$ turns the total number of events into an integral over $y$:
\begin{align}
    N
    &=\int\limits_\textrm{over\,$\mathbf{x}$}~\underbrace{\frac{\partial^n N}{\partial x_1\cdots\partial x_n}}_{Nf(\mathbf{x})}
    ~\underbrace{\mathrm{d}x_1\cdots\mathrm{d}x_n\phantom{\frac{1^n}{2_n}}\!\!\!\!\!\!}_{\mathrm{d}\mathbf{x}}
    \label{eq:N0}\\
    &=\textstyle \int\limits_\textrm{over\,$y$}~\underbrace{\textstyle \int\limits_{\textrm{over\,}\mathbf{x}}\delta(f(\mathbf{x})-y)\,Nf(\mathbf{x})\,\mathrm{d}\mathbf{x}}_{\mathrm{d}{N}/\mathrm{d}{y}}\mathrm{d}y.
    \tag{\ref{eq:N0}$'$}
\end{align}
We express the $\delta$ with the Heaviside step function $\theta$:
\begin{align}
    \frac{1}{N}\frac{\mathrm{d}{N}}{\mathrm{d}{y}}
    =
    y\frac{\mathrm{d}}{\mathrm{d}y}
     \underbrace{\int\theta(f(\mathbf{x})-y)\,\mathrm{d}\mathbf{x}}_\textrm{volume $V$}
    =
    \frac{\mathrm{d}V}{\mathrm{d}(-\ln y)}
    \label{eq:H}
\end{align}
and define the volume of phase space $V{\,=\,}V(y)$.
The integral in Eqn.~\ref{eq:H} picks out the $(n{-}1)$d surface $S$ where $f(\mathbf{x}){\,=\,}y$.
Near $S$ we write $\mathrm{d}\mathbf{x}{\,=\,}\mathrm{d}x_\perp\,\mathrm{d}\mathbf{a}$, with $x_\perp$ along the normal, so that $f(\mathbf{x}){\,-\,}y{\,=\,}|\nabla f|\,x_\perp{\,+\,}\mathcal{O}(x_\perp^2)$.
The integral over $\mathbf{x}$ without the factor $y$ is, to first order, 
\begin{align}
    \!\!\frac{\mathrm{d}V}{\mathrm{d}y}
    {\,=\,}
    \oint\!\!\!\int
    \!\underbrace{\delta\big(|\nabla f|\,x_\perp\big)\vphantom{\Bigg|}
    }_{\delta(x_\perp)/|\nabla f|}
    \,\mathrm{d}x_\perp\mathrm{d}\mathbf{a}
    {\,=\,}
    \!\underbrace{\oint_{f(\mathbf{x})=y}\vphantom{\Bigg|}
    }_{\textrm{surface $S$}}\!\!\!\!\!\!\!\!\big|\nabla f(\mathbf{x})\big|^{-1}\,\mathrm{d}\mathbf{a}.
    \!
    \label{eq:oint}
\end{align}
The first equality uses the $\delta$ property and the absolute value makes the inward/outward orientation irrelevant;
the second integrates over the $x_\perp$.
We have arrived at a volume-surface relation \cite{footnote:Boltzmann}, after restoring the $y$:
\begin{align}
    \frac{1}{N}
    \frac{\mathrm{d}N}{\mathrm{d}y}
    {\,=}
    \underbrace{
    \frac{\mathrm{d}}{\mathrm{d}({-}\ln y)}\int_{f(\mathbf{x})>y}\!\!\!\!\!\!\!\!\!\mathrm{d}\mathbf{x}
    }_{\textrm{information theoretic}}
    =
    y
    \underbrace{
    \oint_{f(\mathbf{x})=y}\!\!\!\!\!\!\!
    \big|\nabla f(\mathbf{x})\big|^{-1}\mathrm{d}\mathbf{a}.
    }_{\textrm{geometric}}
\end{align}
The $V(y)$ is that of the higher density region of $f$ above $y$~\cite{Polonik:1995,Hyndman:1996}, the prior mass of nested sampling~\cite{Skilling:2006} and thresholded to compute the anomaly score~\cite{Scholkopf:2001,Scott:2006,Clemencon:2013}.

As an example, we consider a toy model.
Two samples are created, at $10^6$ events each, of a spin-1 particle decaying to two spin-$\frac{1}{2}$ fermions, with the helicity fractions $F_0{\,=\,}0.7$ and $F_L{\,=\,}0.3$ with respect to a spin axis to mimic a sample of $W$ decays and of a spin-0 particle decaying isotropically.
The contours represent equiprobable lines for constant values of $y$.
The ratio of the two samples $N_0/N_1$ increases, as shown by the red curve, as $y{\,\rightarrow\,}0$.

\begin{figure}[b!]
    \centering
    \includegraphics[width=0.95\linewidth]{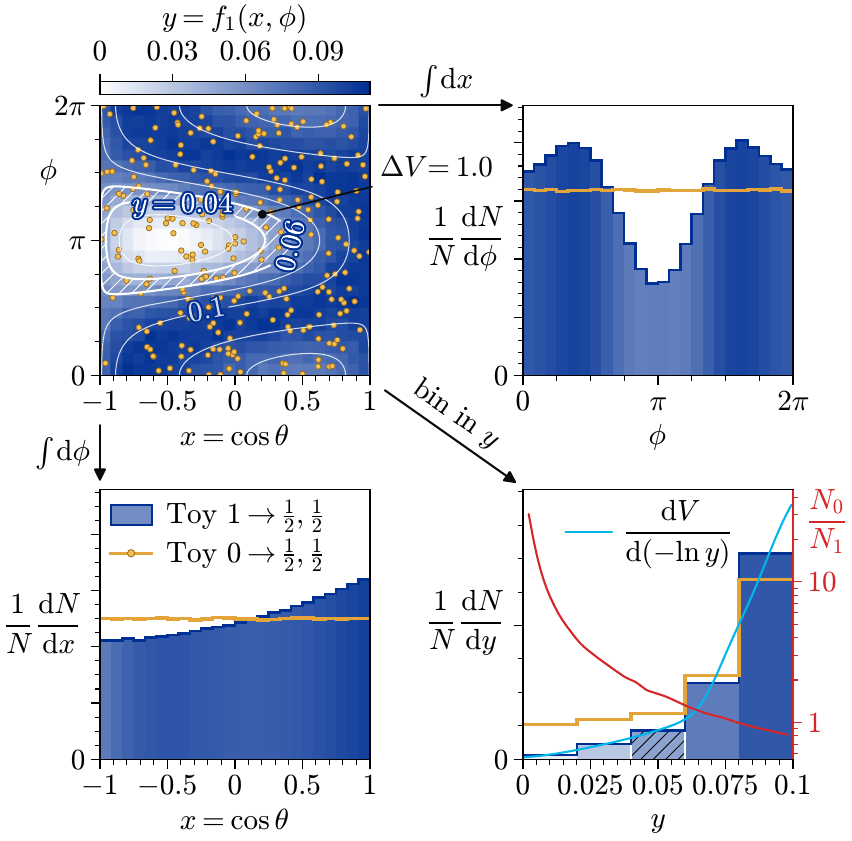}
    \vspace{-9pt}%
    \caption{
    Toy sample of spin-1 decay to fermions to construct $f_1(\cos\theta,\phi)$ and applied to spin-0.
    Plots show $\mathrm{d}^2N/\mathrm{d}\Omega$ and projections.
    The shaded annular region in $\mathrm{d}^2N/\mathrm{d}\Omega$, indicated as $\Delta V$ between $0.04{\,<\,}y{\,<\,}0.06$ corresponds to the shaded bin in $\mathrm{d}N/\mathrm{d}y$, where the computed $\mathrm{d}V/\mathrm{d}(-\ln{y})$ is also overlaid.
    The ratio $N_0/N_1$ is shown on the $\mathrm{d}N/\mathrm{d}y$ plot.
    }
    \label{fig:toy}
\end{figure}

We work out the numerical example for the shaded bin in Fig.\ \ref{fig:toy},
where the top-left plot shows the differential distribution of the angles $\cos\theta$ and $\phi$; the darker (lighter) color shows regions that are more (less) dense in the spin-1 events;
yellow dots represent a spin-0 particle decay to the same final state.
The gradient is approximately constant along the equiprobable path in this example, so
\begin{align}
    \underbrace{\frac{1}{N}
    \frac{\Delta N}{\Delta y}\vphantom{\Bigg|_y\oint_{f(\mathbf{x})=y}}}_\textrm{counts\vphantom{l}}
    =
    \underbrace{-\bar{y}\,\frac{\Delta V}{\Delta y}\Bigg|_y\vphantom{\Bigg|_y\oint_{f(\mathbf{x})=y}}}_\textrm{volume\vphantom{l}}
    &=
    \underbrace{\bar{y}~\Big\langle|\nabla f(\mathbf{x})\big|^{-1}\Big\rangle_{\!y}{\cdot}
    \oint_{f(\mathbf{x})=y}\!\!\!\mathrm{d}\ell\vphantom{\Bigg|_y\oint_{f(\mathbf{x})=y}}}_\textrm{surface\vphantom{l}},
\end{align}
The density vanishes along the spin axis, at $(\cos\theta,\phi){\,=\,}(-0.5,\pi)$, so $V$ is the area outside the light region around it.
For $0.04{\,<\,}y{\,<\,}0.06$, the three parts of Eqn.~\ref{eq:dNdy} can be read off Fig.~\ref{fig:toy}.
The ring has area $\Delta V{\,=\,}1.04$, and $s{\,=\,}{-}\ln y$ falls from 3.22 to 2.81 across it.
Counts: 2.6.
Volume: $\Delta V/\Delta s{\,=\,}1.04/0.405{\,=\,}2.6$.
Surface: $\bar{y}{\,=\,}0.05$, path length of 5.2 and an average gradient of 0.101, yielding $0.05{\,\cdot\,}5.2{/}0.101{\,=\,}2.6$.


\begin{thebibliography}{99}
\bibitem{Note:Boltzmann}
The Boltzmann distribution provides a physical intuition of $s{\,=\,}{-}\ln{y}$, where
$f(x){\,=\,}e^{-E(x)/kT}/Z$ has a surprisal of $s{\,=\,}E$ up to normalization and shift.
Then the derivative, $\mathrm{d}/\mathrm{d}s{\,\rightarrow\,}\mathrm{d}/\mathrm{d}E$, provides a distribution in $E$.
\end{thebibliography}
\end{document}